\documentclass[a4paper,11pt]{article}
\pdfoutput=1 

\usepackage{jcappub} 

\usepackage[T1]{fontenc} 
\usepackage{bm}
\usepackage{graphicx,subfig,color}
\usepackage{amssymb,amsmath}
\usepackage{slashed}
\usepackage{array}
\usepackage{hyperref}

\title{\boldmath Estimating the dark matter halo velocity and surface temperature of some known pulsars due to dark matter capture}

\author[a]{Debashree Sen}
\author[b,1]{Atanu Guha,\note{Corresponding author}}

\affiliation[a]{Center for Extreme Nuclear Matters (CENuM), Korea University, Seoul 02841, Korea}
\affiliation[b]{Department of Physics, Chungnam National University, 99, Daehak-ro, Yuseong-gu, Daejeon-34134, South Korea}

\emailAdd{debashreesen88@gmail.com}
\emailAdd{atanu@cnu.ac.kr}

\abstract{Considering four known pulsars J1906+0746, J1933-6211, J2043+1711 and the Vela pulsar, we study the scenario of dark matter (DM) capture in neutron stars (NSs). For the purpose we choose four well-known relativistic mean field models to obtain the radius corresponding to the observed mass of these pulsars and consequently the scattering cross-section of DM with the different particles of the $\beta$ stable NS matter. The estimated DM-electron scattering cross-section in this work is stringent compared to the current direct detection experimental probe. We then compute the lower limit on the halo velocity of DM for the four pulsars from the knowledge of the upper limit on effective temperature of the individual pulsars. We also extend our work to calculate the value of the effective temperature with the different models using the fitted values of the halo velocity of DM of the four pulsars with respect to their distances from the galactic center. Our findings are consistent with the analysis of the observed data.}

\begin{document}
\maketitle
\flushbottom

\section{Introduction}
\label{Intro}

The lion's share of the total energy budget of the Universe is made up of dark energy (about $70\%$) while the succeeding leading component is dark matter (around $25\%$). Only $5\%$ of the energy content consists of so-called luminous matter or the baryonic matter. This is nowadays an unambiguously evident fact supported by a numerous astrophysical and cosmological observations~\cite{Planck:2015fie,Bertone:2004pz,Aghanim:2018eyx,Bauer:2020zsj}. Over the last few decades several dedicated search methods have been implemented to get an idea about the interaction strength between the dark matter (DM) and standard model (SM) particles. Among the popular experimental search avenues, significant developments have been made for the direct and indirection detection strategies~\cite{Lisanti:2016jxe,Profumo:2013yn}. The most stringent constraints are obtained till date by the leading direct detection experiments like SuperCDMS~\cite{SuperCDMS:2018mne}, XENONnT~\cite{XENON:2022ltv}, PandaX-II~\cite{PandaX-II:2020oim}, DarkSide-50~\cite{DarkSide:2018ppu}, SENSEI~\cite{Crisler:2018gci} and LUX-ZEPLIN~\cite{LZ:2022lsv}. Experiments dedicated to indirect search are FERMI-LAT~\cite{Fermi-LAT:2009ihh}, IceCube~\cite{IceCube:2014stg}, PAMELA~\cite{PAMELA:2008gwm, PAMELA:2013vxg}, AMS-02~\cite{AMS:2014bun}, Voyager~\cite{Boudaud:2016mos} and CALET~\cite{CALET:2017uxd,Adriani:2018ktz}.

Alongside the progress of the experimental probes, there are significant advancement in the phenomenological aspects as well. One of the well motivated methods to understand the properties of the DM particles is to explore the DM capture by neutron stars (NSs) as suggested by the recent literature~\cite{Baryakhtar:2017dbj,Joglekar:2019vzy,Bell:2019pyc,Joglekar:2020liw,Bramante:2023djs}. NSs are the most dense objects present in our Universe which are accessible by direct observations. They serve the purpose of being a unique natural astrophysical laboratory where we can investigate the properties of matter under extreme conditions. Apart from DM capture by NSs, there are several other mechanisms which can be responsible for the possible presence of DM in NSs, e.g., creation of its own DM through dark decays of neutrons~\cite{Husain:2022bxl,Husain:2022brl,Husain:2023fwb,Zhou:2023ndi} or inheritance of DM from the supernovae~\cite{Nelson:2018xtr} etc. For the present work we are interested in the accretion mechanism. Due to strong gravitational pull, halo DM particles fall into NS and become a structural part of it \cite{Sen:2021wev,Guha:2021njn,Guha:2024pnn,Shakeri:2022dwg,Karkevandi:2021ygv,Lenzi:2022ypb,Lourenco:2022fmf}. Through this accretion process DM gets captured by NS which allows us to explore the interaction strength between DM and SM particles. 

The accreted DM particles lose kinetic energy due to collisions with the matter inside the NS and dump the energy to the NS which in turn heat up the NS. This phenomena can manifest itself as the non-zero surface temperature ($T_S$) of sufficiently cold NSs and the mechanism of kinetic heating can be an observable signature of DM. Due to the self interactions, the DM particles attain thermal equilibrium among themselves and become gravitationally bound to the NS~\cite{Bell:2019pyc, Bell:2020lmm}. Recent successful observations by the Chandra and XMM-Newton~\cite{Tananbaum:2014oda} are reliable and provide an upper bound on the effective temperature ($T_{\infty}$) of the NS surface as seen by a distant observer~\cite{Prinz:2015jkd,Prinz:2015}.

We organize the present work as follows. In the next Sec.~\ref{Formalism}, we briefly address the framework of the four Relativistic Mean Field (RMF) hadronic models, viz., TM1 \cite{Sugahara:1993wz}, GM1 \cite{Glendenning:1991es}, DD2 \cite{Typel:2009sy} and DD-MEX \cite{Taninah:2019cku}. In the same section, we also discuss the mechanism of estimating the threshold values for the DM-SM scattering cross-section ($\sigma_{th}^i$) with different constituents of the $\beta$ stable NS matter. Additionally, we formulate the lower limit on the DM halo velocity ($v_{halo}^{ll}$) from the knowledge of the upper bound on $T_{\infty}$~\cite{Prinz:2015jkd,Prinz:2015} and on the other hand we present an analytical form for the $T_{\infty}$ from the known $v_{halo}$ data~\cite{Bhattacharjee:2013exa}. We then present the results of our estimations based on Sec.~\ref{Formalism} and corresponding discussions in Sec.~\ref{Results} for some known pulsars like J1906+0746, J1933-6211, J2043+1711 and the Vela pulsar, whose mass, position and distance with respect to earth are observationally known. In the absence of any concrete observational data for the radius of these four pulsars, the four RMF models are used to determine the radius of these pulsars corresponding to their masses in order to calculate $\sigma_{th}^i$. We summarize and conclude in the final section \ref{Conclusion} of the paper.


\section{Formalism}
\label{Formalism}

\subsection{Models}

We consider four well-known RMF models. Of them, TM1 \cite{Sugahara:1993wz} and GM1 \cite{Glendenning:1991es} are with non-linear self couplings while DD2 \cite{Typel:2009sy} and DD-MEX \cite{Taninah:2019cku} have density-dependent couplings following the Typel-Wolter ansatz \cite{Lu:2011wy}. The values of the couplings and the saturation properties (like the saturation density $\rho_0$, the symmetry energy $J_0$, the slope parameter $L_0$, nuclear incompressibility $K_0$, skewness coefficient $S_0$, and the curvature parameter $K_{sym}$ of the nuclear symmetry energy) of all the above four models considered in this present work, can be found in the respective references and also in \cite{Xia:2022dvw}. Also the detailed methodology for obtaining the equation of state (EoS) of $\beta$ equilibrated NS matter using these four models can be found in the respective references and in \cite{Xia:2022dvw,Guha:2024pnn}. 

For the obtained EoS, we compute the radius ($R_{NS}$) corresponding to the observed mass ($M_{NS}$) of the pulsars J1906+0746, J1933-6211, J2043+1711 and the Vela pulsar, with the help of the Tolman-Oppenheimer-Volkoff (TOV) equations \cite{Tolman:1939jz,Oppenheimer:1939ne}. $M_{NS}$ of J1906+0746, J1933-6211, and J2043+1711 is taken from \url{https://www3.mpifr-bonn.mpg.de/staff/pfreire/NS_masses.html} while that of the Vela pulsar from \cite{Quaintrell:2003pn}. In Tab. \ref{tab:1} we specify the measured quantities like the observed mass ($M_{NS}$), distance from the earth ($d$), right ascension (RA) and declination (Dec) of these four pulsars used to obtain the results of our present work.

\begin{table}
\centering
\caption{Some known pulsars and measured quantities like mass ($M_{NS}$), distance from the earth ($d$), right ascension (RA) and declination (Dec). $M_{NS}$ of J1906+0746, J1933-6211, and J2043+1711 is taken from \url{https://www3.mpifr-bonn.mpg.de/staff/pfreire/NS_masses.html} and that of the Vela pulsar from \protect\cite{Quaintrell:2003pn}}
\setlength{\tabcolsep}{5.0pt}
\begin{tabular}{cccccc}
\hline
\hline
Pulsar & $M_{NS}$ & $d$ & RA & Dec \\
& ($M_{\odot}$) & (kpc) &  &  \\
\hline
J1906+0746 & 1.29(11)           & 7.4   & 19h 06m 48.86s &  07$^{\circ}$ 46' 25.9" \\
J1933-6211 & 1.4$^{+3}_{-2}$    & 0.65  & 19h 33m 32.43s & -62$^{\circ}$ 11' 46.88" \\
J2043+1711 & 1.38$^{+12}_{-13}$ & 1.389 & 20h 43m 20.88s &  17$^{\circ}$ 11' 28.89" \\
Vela       & 1.88$\pm$0.13      & 0.28  & 08h 35m 20.61s & -45$^{\circ}$ 10' 34.87" \\
\hline
\hline
\end{tabular}
\label{tab:1}
\end{table} 
 
\subsection{Dark matter capture and threshold scattering cross-section}

The threshold scattering cross-section ($\sigma_{th}^i$) of the DM with the $i-$th particle can be obtained following \cite{Bell:2019pyc} as
\begin{eqnarray}
\sigma_{th}^i = \frac{\pi R_{NS}^2}{N_i} \frac{k_i}{q} , &{\rm{if}}~ q < k_i \nonumber\\
&{\rm{otherwise}}~ k_i/q=1
\label{sigma_th}
\end{eqnarray}
where, $k_i$ is the Fermi momentum of the $i-$th species. Since we consider $\beta$ equilibrated NS matter, $i=n,p,e,\mu$ and $N_i$ is the total individual particle density up to $R_{NS}$, obtained by radially integrating the density profile $N_i(r)$ of each particle obtained at different values of the radius up to $R_{NS}$. The momentum transfer $q$ is given as
\begin{eqnarray}
q=(\gamma_{esc}-1)m_{\chi} v_{\chi}
\end{eqnarray}
where, $m_{\chi}$ is the mass of the captured fermionic DM and $v_{\chi}$ its velocity. We have 
\begin{eqnarray}
v_{\chi}=v_{esc}=\frac{\sqrt{\gamma_{esc}^2 - 1}}{\gamma_{esc}}
\end{eqnarray}
because we have considered maximum impact for incident DM following \cite{Joglekar:2020liw} and
\begin{eqnarray}
\gamma_{esc}=1+\frac{2M_{NS}}{R_{NS}}
\label{gesc}
\end{eqnarray}

\subsection{Lower limit on $v_{halo}$}

We follow \cite{Bell:2019pyc,Joglekar:2020liw} to obtain the lower limit on $v_{halo}$ i.e. the velocity of DM at NS halo before capture. The surface temperature $T_S$ of NS can be obtained in terms of the upper limit on effective temperature $T_{\infty}$ \cite{Yakovlev:2004iq} as
\begin{eqnarray}
T_S=\frac{T_{\infty}}{\sqrt{1-\frac{2M_{NS}}{R_{NS}}}}
\label{Tsurf}
\end{eqnarray} 
Adopting the upper limit of $T_{\infty}$ for the four pulsars considered from \cite{Prinz:2015jkd,Prinz:2015}, we obtain the upper limit on $T_S$ and consequently on the energy deposited per unit time by DM ($\dot{k}$) using the expression
\begin{eqnarray}
\dot{k}\bigg(1-\frac{2M_{NS}}{R_{NS}}\bigg)=4\pi R_{NS}^2 \sigma_{SB} T_S^4
\label{kdot}
\end{eqnarray}
where, the Stefan-Boltzman constant $\sigma_{SB}$=$\pi^2/60$.
The kinetic energy at the surface when the DM enters the NS is given as 
\begin{eqnarray}
E_S=(\gamma_{esc}-1)m_{\chi}
\end{eqnarray}
Therefore we get
\begin{eqnarray}
\dot{k}=E_S \frac{\dot{M}_{\chi}}{m_{\chi}} f
\label{kdot2}
\end{eqnarray}
where, the DM mass passing through surface of NS per unit time ($\dot{M_{\chi}}$) is given as 
\begin{eqnarray}
\dot{M}_{\chi}=\pi b_{max}^2 v_{esc} \rho_{\chi}
\end{eqnarray}
Here, the maximum impact parameter for the DM to intersect the NS surface is
\begin{eqnarray}
b_{max}=R_{NS} \frac{v_{esc}}{v_{halo}} \gamma_{esc}
\label{bmax}
\end{eqnarray}
and $\rho_{\chi}$ is the mass density of DM halo.
In Eq. \ref{kdot2}, the capture fraction
\begin{eqnarray}
f=min\left[\frac{\sigma_{\chi}^i}{\sigma_{th}^i}, 1 \right]
\label{f}
\end{eqnarray}
Here, $f=$1 because we have considered the upper bound on $T_{\infty}$ which always makes $f=$1. In \cite{Bell:2019pyc} it is already mentioned that if $T_{\infty} >$ 1700 Kelvin (NS blackbody temperature), the DM capture rate is maximum i.e., $f=$1. 

Combining Eqs. \ref{kdot} - \ref{bmax} we have
\begin{eqnarray}
\frac{\rho_{\chi}}{v_{halo}^2}=\frac{\dot{k}}{\left(\gamma_{esc} - 1\right) \pi R_{NS}^2 \left(\gamma_{esc}^2 - 1\right)^{3/2}} \gamma_{esc}
\label{vhalo}
\end{eqnarray}
where, $\gamma_{esc}$ is obtained from Eq. \ref{gesc} in terms of $M_{NS}$ and $R_{NS}$. As we have the upper limit on $\dot{k}$ using the upper limit of $T_{\infty}$, we obtain lower limit on $v_{halo}$ as $v_{halo}^{ll}$ from Eq. \ref{vhalo}. Corresponding to the mass $M_{NS}$ of each pulsar, we have four values of $R_{NS}$ with four different RMF models TM1, GM1, DD2 and DD-MEX. Thus we have four slightly different upper limit of $T_{\infty}$ for each pulsar.
In Eq. \ref{vhalo} the $\rho_{\chi}$ for the four pulsars is obtained using the Navarro-Frenk-White (NFW) profile \cite{Navarro:1995iw,Bauer:2017qwy} as
\begin{eqnarray}
\rho_{\chi}\left(d_{GC}\right)=\frac{\rho_0}{\left(\frac{d_{GC}}{r_s}\right)\left({1+\frac{d_{GC}}{r_s}}\right)^2}
\end{eqnarray}
where, $d_{GC}$ is the distance of the particular pulsar from galactic center, calculated in Appendix \ref{Appendix} using the values of the distance of pulsar from earth ($d$) and its RA and Dec. $\rho_0$ and $r_s$ are the reference parameters \cite{Navarro:1995iw,Bauer:2017qwy}.

\subsection{Effective temperature from $v_{halo}$ data}

We next calculate the surface temperature of the four individual pulsar considering the fitted values of $v_{halo}$ data reported in \cite{Bhattacharjee:2013exa} with respect to $d_{GC}$. Once we obtain the values of $\rho_{\chi}(d_{GC})$ for the individual pulsars and the values of $v_{halo}$ from the fitted data corresponding to $d_{GC}$ referring \cite{Bhattacharjee:2013exa}, we compute $\dot{k}$ using Eq. \ref{vhalo} and consequently we obtain the values of surface temperature $T_S$ from Eq. \ref{kdot}. Finally we compute the effective temperature $T_{\infty}$ of each pulsar using Eq. \ref{Tsurf}. With four different RMF models, we have four slightly different values of $T_{\infty}$ for each pulsar.


\section{Results and Discussions}
\label{Results}

\subsection{Dark matter capture and lower limit on $v_{halo}$}
\label{sec:vhalo}

\begin{table}[!ht]
\centering
\caption{Some known pulsars and their effective temperature ($T_{\infty}^{ul}$) taken from \protect\cite{Prinz:2015jkd,Prinz:2015} and the obtained radius ($R_{NS}$), distance from galactic center ($d_{GC}$), and the lower limit on $v_{halo}$ ($v_{halo}^{ll}$) with the four different RMF models.}
\setlength{\tabcolsep}{5.0pt}
\begin{tabular}{cccccc}
\hline
\hline
Pulsar & $d_{GC}$ & $\log_{10}T_{\infty}$ & Model  & $R_{NS}$ & $v_{halo}^{ll}$ \\
& (kpc) & & & (km) & \\
\hline
J1906+0746 & 5.55   & 5.93   & TM1     & 12.47 & 3.495$\times$10$^{-9}$  \\
           &        &        & GM1     & 12.23 & 3.534$\times$10$^{-9}$  \\
           &        &        & DD2     & 11.99 & 3.572$\times$10$^{-9}$  \\
           &        &        & DD-MEX  & 12.85 & 3.434$\times$10$^{-9}$  \\
\hline
J1933-6211 & 7.595  & 5.64   & TM1     & 12.63 & 1.073$\times$10$^{-8}$  \\
           &        &        & GM1     & 12.37 & 1.084$\times$10$^{-8}$  \\
           &        &        & DD2     & 12.15 & 1.094$\times$10$^{-8}$  \\
           &        &        & DD-MEX  & 12.98 & 1.058$\times$10$^{-8}$  \\
\hline
J2043+1711 & 7.572  & 5.91   & TM1     & 12.60 & 3.085$\times$10$^{-9}$  \\
           &        &        & GM1     & 12.35 & 3.116$\times$10$^{-9}$  \\
           &        &        & DD2     & 12.13 & 3.143$\times$10$^{-9}$  \\
           &        &        & DD-MEX  & 12.96 & 3.039$\times$10$^{-9}$  \\
\hline
Vela       & 8.14   & 5.82   & TM1     & 12.93 & 4.833$\times$10$^{-9}$  \\
           &        &        & GM1     & 12.65 & 4.841$\times$10$^{-9}$  \\
           &        &        & DD2     & 12.52 & 4.844$\times$10$^{-9}$  \\
           &        &        & DD-MEX  & 13.30 & 4.815$\times$10$^{-9}$  \\
\hline
\hline
\end{tabular}
\label{tab:2}
\end{table}

In Tab. \ref{tab:2} we display the upper limit effective temperature ($T_{\infty}^{ul}$) taken from \cite{Prinz:2015jkd,Prinz:2015} of the four pulsars and the obtained radius ($R_{NS}$), distance from galactic center ($d_{GC}$) according to Appendix \ref{Appendix}, and the lower limit on $v_{halo}$ ($v_{halo}^{ll}$) with the four different RMF models. We find that for any particular pulsar, the value of $v_{halo}^{ll}$ does not differ very much for the different models. From Tab. \ref{tab:2}, it is clear that for any given pulsar of a particular mass, the value of $v_{halo}^{ll}$ is minimum for the model that gives the maximum radius. Our values of $v_{halo}^{ll}$ are also consistent with i.e., less than the fitted values of $v_{halo}$ obtained in \cite{Bhattacharjee:2013exa} with respect to the $d_{GC}$ of each pulsar.

\subsection{Effective temperature from $v_{halo}$ data}
\label{Tinf}

\begin{table}[!ht]
\centering
\caption{Some known pulsars and their values of $v_{halo}$ fitted from \protect\cite{Bhattacharjee:2013exa} and the obtained effective temperature ($T_{\infty}$) with the four different RMF models.}
\setlength{\tabcolsep}{5.0pt}
\begin{tabular}{cccccc}
\hline
\hline
Pulsar & $v_{halo}$ & Model & $\log_{10}T_{\infty}$  \\
\hline
J1906+0746 & 8.452$\times$10$^{-4}$  & TM1     & 3.989  \\
           &                         & GM1     & 3.991  \\
           &                         & DD2     & 3.993  \\
           &                         & DD-MEX  & 3.985  \\
\hline
J1933-6211 & 8.931$\times$10$^{-4}$  & TM1     & 3.928  \\
           &                         & GM1     & 3.930  \\
           &                         & DD2     & 3.932  \\
           &                         & DD-MEX  & 3.925  \\
\hline
J2043+1711 & 8.928$\times$10$^{-4}$  & TM1     & 3.927  \\
           &                         & GM1     & 3.930  \\
           &                         & DD2     & 3.931  \\
           &                         & DD-MEX  & 3.924  \\
\hline
Vela       & 8.994$\times$10$^{-4}$  & TM1     & 3.935  \\
           &                         & GM1     & 3.935  \\
           &                         & DD2     & 3.935  \\
           &                         & DD-MEX  & 3.925  \\
\hline
\hline
\end{tabular}
\label{tab:3}
\end{table}

In Tab. \ref{tab:3} we tabulate the values of $T_{\infty}$ obtained for the four pulsars with the different models using the fitted values of $v_{halo}$ with respect to $d_{GC}$ from \cite{Bhattacharjee:2013exa}. The obtained values of $T_{\infty}$ in Tab. \ref{tab:3} are consistent with i.e., less than the upper limit on $T_{\infty}$ prescribed in \cite{Prinz:2015jkd,Prinz:2015}. It can be seen from Tab. \ref{tab:3} that for any particular pulsar, the value of $T_{\infty}$ differs negligibly for different models.

\newpage

\subsection{Dark matter scattering cross-section}

\begin{figure}[!ht]
\centering
\subfloat[J1906+0746]{\includegraphics[width=0.49\textwidth]{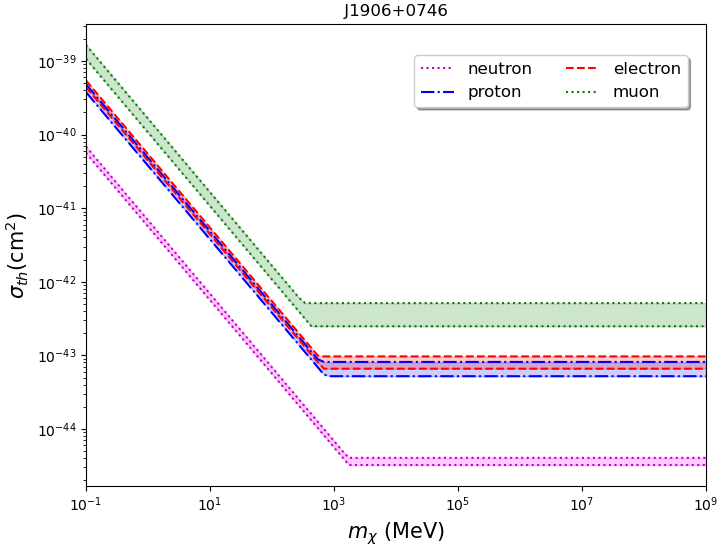}\protect\label{J1906}}
\subfloat[J1933-6211]{\includegraphics[width=0.49\textwidth]{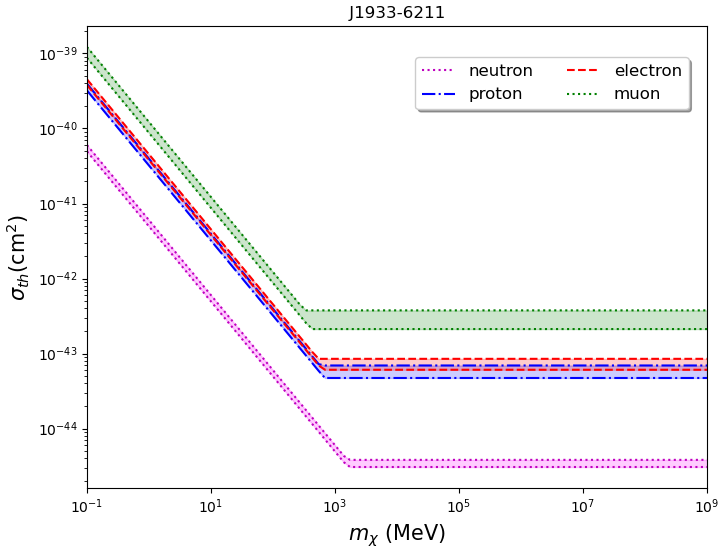}\protect\label{J1933}}
\hfill
\subfloat[J2043+1711]{\includegraphics[width=0.49\textwidth]{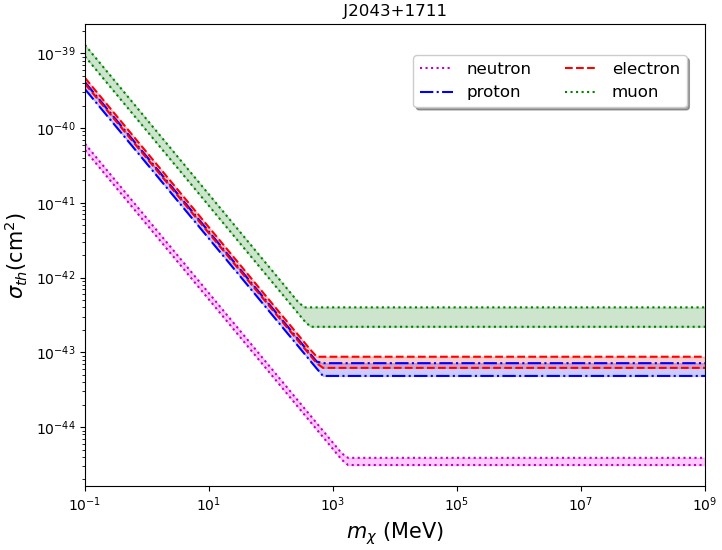}\protect\label{J2043}}
\subfloat[Vela]{\includegraphics[width=0.49\textwidth]
{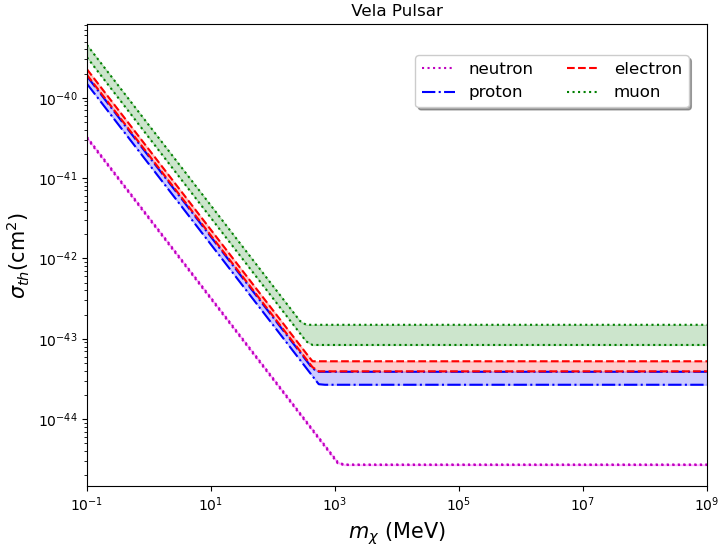}\protect\label{vela}}
\caption{\it Threshold scattering cross-section of dark matter with respect to mass of dark matter with neutrons, protons, electrons and muons obtained as range with the four models for the four pulsars.}
\label{sigma}
\end{figure}

In Fig. \ref{sigma} we show the variation of threshold scattering cross-section of DM with the different particle species (viz., neutrons, protons, electrons and muons) with respect to mass of fermionic DM using the four models for the four individual pulsars. The results obtained with the four RMF models are displayed as bands in Figs. \ref{J1906}, \ref{J1933}, \ref{J2043} and \ref{vela} for the pulsars J1906+0746, J1933-6211, J2043+1711 and the Vela pulsar, respectively. The value of threshold cross-section $\sigma_{th}^i$ decreases initially when $q < k_i$ and then attains a constant value when $q = k_i$ (recalling Eq. \ref{sigma_th}). When $q < k_i$, the order of $\sigma_{th}^i$ is almost same for J1906+0746, J1933-6211, and J2043+1711 ($\sim$10$^{-39}$) and slightly higher than 10$^{-40}$ in case of the Vela pulsar. We also find that the cross-section is minimum for neutrons while it is maximum for muons. This is because the muons populate the NS matter the least and hence $N_{\mu}$ is minimum compared to the other constituents while $N_n$ is maximum. Therefore, from Eq. \ref{sigma_th} it can be easily understood that $\sigma_{th}^{\mu}$ is maximum while $\sigma_{th}^n$ minimum for all the pulsars. In Fig. \ref{sigma} the bands signify the results from different RMF models. However, in case of Vela pulsar we obtain very negligible difference in the results of $\sigma_{th}^n$ with the four models as seen from Fig. \ref{vela}. The values $\sigma_{th}^n$ thus appear to be almost overlapping in Fig. \ref{vela}.

\begin{figure}[!ht]
\centering
{\includegraphics[width=0.5\textwidth]{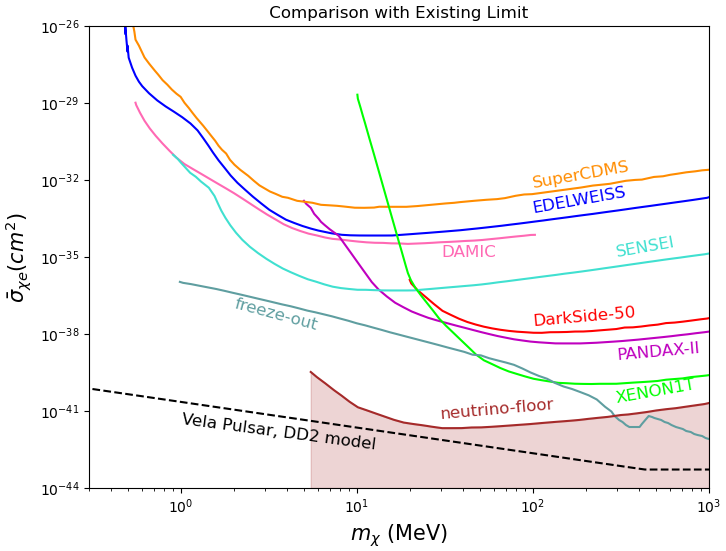}}
\caption{\it Actual scattering cross-section of dark matter with respect to mass of dark matter with electrons obtained with the DD2 model for the Vela pulsar. The existing bounds from different direct detection experiments are also compared which are taken from SENSEI~\protect\cite{SENSEI:2020dpa}, EDELWEISS~\protect\cite{EDELWEISS:2020fxc}, PandaX-II~\protect\cite{PandaX-II:2021nsg}, XENON1T~\protect\cite{XENON:2019gfn}, SuperCDMS~\protect\cite{SuperCDMS:2020ymb}, DAMIC~\protect\cite{DAMIC:2019dcn}, and DarkSide-50~\protect\cite{DarkSide:2018ppu}.}
\label{sigma_existing}
\end{figure} 

As discussed in Sec. \ref{sec:vhalo}, the capture rate is maximum. Hence the actual cross-section is always equal to the threshold cross-section i.e, $\bar{\sigma}_{\chi i}$=$\sigma_{th}^i$. In Fig. \ref{sigma_existing} we show the DM-electron scattering cross-section $\bar{\sigma}_{\chi e}$. Since the results of cross-section are almost same for the four different models, we choose to show the one obtained with DD2 model for the Vela pulsar in Fig. \ref{sigma_existing}. We also compare the existing bounds from different direct detection experiments, taken from SENSEI~\cite{SENSEI:2020dpa}, EDELWEISS~\cite{EDELWEISS:2020fxc}, PandaX-II~\cite{PandaX-II:2021nsg}, XENON1T~\cite{XENON:2019gfn}, SuperCDMS~\cite{SuperCDMS:2020ymb}, DAMIC~\cite{DAMIC:2019dcn}, and DarkSide-50~\cite{DarkSide:2018ppu}. It is seen that our results are well beyond the explored region of the parameter space. Our estimates of $\bar{\sigma}_{\chi e}$ is also quite consistent with that obtained in \cite{Bell:2019pyc}.


\section{Summary and Conclusion}
\label{Conclusion}

The present work revolves around the context of the kinetic heating of some known pulsars like J1906+0746, J1933-6211, J2043+1711 and the Vela pulsar due to the DM capture by them. For these pulsars, the mass, position with respect to the celestial sphere (RA and Dec) and distance from the earth are precisely measured to some extent. We utilize the EoS obtained with the well known RMF models (TM1, GM1, DD2 and DD-MEX) as the input to the TOV equations and obtain the radius of these four pulsars. The prediction of the radius and the radial profile of the number densities are a bit different for the different models considered. Therefore, for each pulsar, we obtain the threshold values of the interaction cross-section of DM with individual SM particles in NS to be of the same order for the different models. We compare the strongest scattering cross-section for the DM-electron interaction which is obtained for the Vela pulsar along with the experimentally explored region of the parameter space. Our prediction of the DM-electron scattering cross-section is a few order smaller than the values probed by the leading direct detection experiments and therefore this can act as a guide to the future developments of the direct and indirect searches of DM. 

We extend our work to compute the lower bound on the halo velocity of DM from the knowledge of the upper bound on the effective temperature of the pulsars~\cite{Prinz:2015jkd,Prinz:2015}. This lower bound on $v_{halo}$ is consistent with the results obtained from the analysis of the observational data \cite{Bhattacharjee:2013exa}.

On a different note we also estimate the value of $T_\infty$ utilizing the observational findings of $v_{halo}$. Our estimates are consistent with the upper bound on $T_\infty$ \cite{Prinz:2015jkd,Prinz:2015}.

\section*{Acknowledgements}

Work of DS is supported by the NRF research Grants (No. 2018R1A5A1025563). Work of A.G. is supported by the National Research Foundation of Korea (NRF-2019R1C1C1005073).


\appendix

\section{Distance of pulsars from galactic center}
\label{Appendix}

We obtain the distance $d_{GC}$ of the four pulsars from galactic center using their individual RA and Dec and distance from earth ($d=r_2$). For the galactic center we also know that RA = 17h 45m 40.4s and Dec = 17$^{\circ}$ 00' 28.11" and the distance of the galactic center from the earth is $r_1$=8.1 kpc. Using the fact that 24h=2$\pi$, we can convert the RA's into radian as $\alpha$ and also using 1$^{\circ}$=$\frac{\pi}{180}$, the Dec's can also be converted to radian as $\beta$.

Now in spherical polar coordinates, we have $\alpha=\phi$ and $\theta=\frac{\pi}{2} - \beta$. Thus the distance of any pulsar from galactic center is given as
\begin{eqnarray}
d_{GC}=\sqrt{r_1^2 + r_2^2 - 2r_1r_2[sin\theta_1 sin\theta_2cos(\phi_1 - \phi_2) + cos\theta_1 cos\theta_2]}
\end{eqnarray} 
where, ($r_1$, $\theta_1$, $\phi_1$) is the coordinate of the galactic center and ($r_2$, $\theta_2$, $\phi_2$) is that of a pulsar.

\bibliography{ref}  

\providecommand{\href}[2]{#2}\begingroup\raggedright\begin{thebibliography}{10}

\bibitem{Planck:2015fie}
{\scshape Planck} collaboration, \emph{{Planck 2015 results. XIII. Cosmological
  parameters}},
  \href{https://doi.org/10.1051/0004-6361/201525830}{\emph{Astron. Astrophys.}
  {\bfseries 594} (2016) A13}
  [\href{https://arxiv.org/abs/1502.01589}{{\ttfamily 1502.01589}}].

\bibitem{Bertone:2004pz}
G.~Bertone, D.~Hooper and J.~Silk, \emph{{Particle dark matter: Evidence,
  candidates and constraints}},
  \href{https://doi.org/10.1016/j.physrep.2004.08.031}{\emph{Phys. Rept.}
  {\bfseries 405} (2005) 279}
  [\href{https://arxiv.org/abs/hep-ph/0404175}{{\ttfamily hep-ph/0404175}}].

\bibitem{Aghanim:2018eyx}
{\scshape Planck} collaboration, \emph{{Planck 2018 results. VI. Cosmological
  parameters}},
  \href{https://doi.org/10.1051/0004-6361/201833910}{\emph{Astron. Astrophys.}
  {\bfseries 641} (2020) A6}
  [\href{https://arxiv.org/abs/1807.06209}{{\ttfamily 1807.06209}}].

\bibitem{Bauer:2020zsj}
J.B.~Bauer, D.J.E.~Marsh, R.~Hlo\v{z}ek, H.~Padmanabhan and A.~Lagu\"e,
  \emph{{Intensity Mapping as a Probe of Axion Dark Matter}},
  \href{https://doi.org/10.1093/mnras/staa3300}{\emph{Mon. Not. Roy. Astron.
  Soc.} {\bfseries 500} (2020) 3162}
  [\href{https://arxiv.org/abs/2003.09655}{{\ttfamily 2003.09655}}].

\bibitem{Lisanti:2016jxe}
M.~Lisanti, \emph{{Lectures on Dark Matter Physics}},  in \emph{{Theoretical
  Advanced Study Institute in Elementary Particle Physics}: {New Frontiers in
  Fields and Strings}}, pp.~399--446, 2017,
  \href{https://doi.org/10.1142/9789813149441_0007}{DOI}
  [\href{https://arxiv.org/abs/1603.03797}{{\ttfamily 1603.03797}}].

\bibitem{Profumo:2013yn}
S.~Profumo, \emph{{Astrophysical Probes of Dark Matter}},  in
  \emph{{Theoretical Advanced Study Institute in Elementary Particle Physics}:
  {Searching for New Physics at Small and Large Scales}}, pp.~143--189, 2013,
  \href{https://doi.org/10.1142/9789814525220_0004}{DOI}
  [\href{https://arxiv.org/abs/1301.0952}{{\ttfamily 1301.0952}}].

\bibitem{SuperCDMS:2018mne}
{\scshape SuperCDMS} collaboration, \emph{{First Dark Matter Constraints from a
  SuperCDMS Single-Charge Sensitive Detector}},
  \href{https://doi.org/10.1103/PhysRevLett.121.051301}{\emph{Phys. Rev. Lett.}
  {\bfseries 121} (2018) 051301}
  [\href{https://arxiv.org/abs/1804.10697}{{\ttfamily 1804.10697}}].

\bibitem{XENON:2022ltv}
{\scshape XENON} collaboration, \emph{{Search for New Physics in Electronic
  Recoil Data from XENONnT}},
  \href{https://doi.org/10.1103/PhysRevLett.129.161805}{\emph{Phys. Rev. Lett.}
  {\bfseries 129} (2022) 161805}
  [\href{https://arxiv.org/abs/2207.11330}{{\ttfamily 2207.11330}}].

\bibitem{PandaX-II:2020oim}
{\scshape PandaX-II} collaboration, \emph{{Results of dark matter search using
  the full PandaX-II exposure}},
  \href{https://doi.org/10.1088/1674-1137/abb658}{\emph{Chin. Phys. C}
  {\bfseries 44} (2020) 125001}
  [\href{https://arxiv.org/abs/2007.15469}{{\ttfamily 2007.15469}}].

\bibitem{DarkSide:2018ppu}
{\scshape DarkSide} collaboration, \emph{{Constraints on Sub-GeV
  Dark-Matter\textendash{}Electron Scattering from the DarkSide-50
  Experiment}},
  \href{https://doi.org/10.1103/PhysRevLett.121.111303}{\emph{Phys. Rev. Lett.}
  {\bfseries 121} (2018) 111303}
  [\href{https://arxiv.org/abs/1802.06998}{{\ttfamily 1802.06998}}].

\bibitem{Crisler:2018gci}
{\scshape SENSEI} collaboration, \emph{{SENSEI: First Direct-Detection
  Constraints on sub-GeV Dark Matter from a Surface Run}},
  \href{https://doi.org/10.1103/PhysRevLett.121.061803}{\emph{Phys. Rev. Lett.}
  {\bfseries 121} (2018) 061803}
  [\href{https://arxiv.org/abs/1804.00088}{{\ttfamily 1804.00088}}].

\bibitem{LZ:2022lsv}
{\scshape LZ} collaboration, \emph{{First Dark Matter Search Results from the
  LUX-ZEPLIN (LZ) Experiment}},
  \href{https://doi.org/10.1103/PhysRevLett.131.041002}{\emph{Phys. Rev. Lett.}
  {\bfseries 131} (2023) 041002}
  [\href{https://arxiv.org/abs/2207.03764}{{\ttfamily 2207.03764}}].

\bibitem{Fermi-LAT:2009ihh}
{\scshape Fermi-LAT} collaboration, \emph{{The Large Area Telescope on the
  Fermi Gamma-ray Space Telescope Mission}},
  \href{https://doi.org/10.1088/0004-637X/697/2/1071}{\emph{Astrophys. J.}
  {\bfseries 697} (2009) 1071}
  [\href{https://arxiv.org/abs/0902.1089}{{\ttfamily 0902.1089}}].

\bibitem{IceCube:2014stg}
{\scshape IceCube} collaboration, \emph{{Observation of High-Energy
  Astrophysical Neutrinos in Three Years of IceCube Data}},
  \href{https://doi.org/10.1103/PhysRevLett.113.101101}{\emph{Phys. Rev. Lett.}
  {\bfseries 113} (2014) 101101}
  [\href{https://arxiv.org/abs/1405.5303}{{\ttfamily 1405.5303}}].

\bibitem{PAMELA:2008gwm}
{\scshape PAMELA} collaboration, \emph{{An anomalous positron abundance in
  cosmic rays with energies 1.5-100 GeV}},
  \href{https://doi.org/10.1038/nature07942}{\emph{Nature} {\bfseries 458}
  (2009) 607} [\href{https://arxiv.org/abs/0810.4995}{{\ttfamily 0810.4995}}].

\bibitem{PAMELA:2013vxg}
{\scshape PAMELA} collaboration, \emph{{Cosmic-Ray Positron Energy Spectrum
  Measured by PAMELA}},
  \href{https://doi.org/10.1103/PhysRevLett.111.081102}{\emph{Phys. Rev. Lett.}
  {\bfseries 111} (2013) 081102}
  [\href{https://arxiv.org/abs/1308.0133}{{\ttfamily 1308.0133}}].

\bibitem{AMS:2014bun}
{\scshape AMS} collaboration, \emph{{High Statistics Measurement of the
  Positron Fraction in Primary Cosmic Rays of 0.5\textendash{}500 GeV with the
  Alpha Magnetic Spectrometer on the International Space Station}},
  \href{https://doi.org/10.1103/PhysRevLett.113.121101}{\emph{Phys. Rev. Lett.}
  {\bfseries 113} (2014) 121101}.

\bibitem{Boudaud:2016mos}
M.~Boudaud, J.~Lavalle and P.~Salati, \emph{{Novel cosmic-ray electron and
  positron constraints on MeV dark matter particles}},
  \href{https://doi.org/10.1103/PhysRevLett.119.021103}{\emph{Phys. Rev. Lett.}
  {\bfseries 119} (2017) 021103}
  [\href{https://arxiv.org/abs/1612.07698}{{\ttfamily 1612.07698}}].

\bibitem{CALET:2017uxd}
{\scshape CALET} collaboration, \emph{{Energy Spectrum of Cosmic-Ray Electron
  and Positron from 10 GeV to 3 TeV Observed with the Calorimetric Electron
  Telescope on the International Space Station}},
  \href{https://doi.org/10.1103/PhysRevLett.119.181101}{\emph{Phys. Rev. Lett.}
  {\bfseries 119} (2017) 181101}
  [\href{https://arxiv.org/abs/1712.01711}{{\ttfamily 1712.01711}}].

\bibitem{Adriani:2018ktz}
O.~Adriani et~al., \emph{{Extended Measurement of the Cosmic-Ray Electron and
  Positron Spectrum from 11 GeV to 4.8 TeV with the Calorimetric Electron
  Telescope on the International Space Station}},
  \href{https://doi.org/10.1103/PhysRevLett.120.261102}{\emph{Phys. Rev. Lett.}
  {\bfseries 120} (2018) 261102}
  [\href{https://arxiv.org/abs/1806.09728}{{\ttfamily 1806.09728}}].

\bibitem{Baryakhtar:2017dbj}
M.~Baryakhtar, J.~Bramante, S.W.~Li, T.~Linden and N.~Raj, \emph{{Dark Kinetic
  Heating of Neutron Stars and An Infrared Window On WIMPs, SIMPs, and Pure
  Higgsinos}},
  \href{https://doi.org/10.1103/PhysRevLett.119.131801}{\emph{Phys. Rev. Lett.}
  {\bfseries 119} (2017) 131801}
  [\href{https://arxiv.org/abs/1704.01577}{{\ttfamily 1704.01577}}].

\bibitem{Joglekar:2019vzy}
A.~Joglekar, N.~Raj, P.~Tanedo and H.-B.~Yu, \emph{{Relativistic capture of
  dark matter by electrons in neutron stars}},
  \href{https://doi.org/10.1016/j.physletb.2020.135767}{\emph{Phys. Lett. B}
  {\bfseries 809} (2020) 135767}
  [\href{https://arxiv.org/abs/1911.13293}{{\ttfamily 1911.13293}}].

\bibitem{Bell:2019pyc}
N.F.~Bell, G.~Busoni and S.~Robles, \emph{{Capture of Leptophilic Dark Matter
  in Neutron Stars}},
  \href{https://doi.org/10.1088/1475-7516/2019/06/054}{\emph{JCAP} {\bfseries
  06} (2019) 054} [\href{https://arxiv.org/abs/1904.09803}{{\ttfamily
  1904.09803}}].

\bibitem{Joglekar:2020liw}
A.~Joglekar, N.~Raj, P.~Tanedo and H.-B.~Yu, \emph{{Dark kinetic heating of
  neutron stars from contact interactions with relativistic targets}},
  \href{https://doi.org/10.1103/PhysRevD.102.123002}{\emph{Phys. Rev. D}
  {\bfseries 102} (2020) 123002}
  [\href{https://arxiv.org/abs/2004.09539}{{\ttfamily 2004.09539}}].

\bibitem{Bramante:2023djs}
J.~Bramante and N.~Raj, \emph{{Dark matter in compact stars}},
  \href{https://doi.org/10.1016/j.physrep.2023.12.001}{\emph{Phys. Rept.}
  {\bfseries 1052} (2024) 1}
  [\href{https://arxiv.org/abs/2307.14435}{{\ttfamily 2307.14435}}].

\bibitem{Husain:2022bxl}
W.~Husain, T.F.~Motta and A.W.~Thomas, \emph{{Consequences of neutron decay
  inside neutron stars}},
  \href{https://doi.org/10.1088/1475-7516/2022/10/028}{\emph{JCAP} {\bfseries
  10} (2022) 028} [\href{https://arxiv.org/abs/2203.02758}{{\ttfamily
  2203.02758}}].

\bibitem{Husain:2022brl}
W.~Husain and A.W.~Thomas, \emph{{Novel neutron decay mode inside neutron
  stars}}, \href{https://doi.org/10.1088/1361-6471/aca1d5}{\emph{J. Phys. G}
  {\bfseries 50} (2023) 015202}
  [\href{https://arxiv.org/abs/2206.11262}{{\ttfamily 2206.11262}}].

\bibitem{Husain:2023fwb}
W.~Husain, D.~Sengupta and A.W.~Thomas, \emph{{Constraining Dark Boson Decay
  Using Neutron Stars}},
  \href{https://doi.org/10.3390/universe9070307}{\emph{Universe} {\bfseries 9}
  (2023) 307} [\href{https://arxiv.org/abs/2306.07509}{{\ttfamily
  2306.07509}}].

\bibitem{Zhou:2023ndi}
D.~Zhou, \emph{{Neutron Star Constraints on Neutron Dark Decays}},
  \href{https://doi.org/10.3390/universe9110484}{\emph{Universe} {\bfseries 9}
  (2023) 484}.

\bibitem{Nelson:2018xtr}
A.~Nelson, S.~Reddy and D.~Zhou, \emph{{Dark halos around neutron stars and
  gravitational waves}},
  \href{https://doi.org/10.1088/1475-7516/2019/07/012}{\emph{JCAP} {\bfseries
  07} (2019) 012} [\href{https://arxiv.org/abs/1803.03266}{{\ttfamily
  1803.03266}}].

\bibitem{Sen:2021wev}
D.~Sen and A.~Guha, \emph{{Implications of feebly interacting dark sector on
  neutron star properties and constraints from GW170817}},
  \href{https://doi.org/10.1093/mnras/stab1056}{\emph{Mon. Not. Roy. Astron.
  Soc.} {\bfseries 504} (2021) 3354}
  [\href{https://arxiv.org/abs/2104.06141}{{\ttfamily 2104.06141}}].

\bibitem{Guha:2021njn}
A.~Guha and D.~Sen, \emph{{Feeble DM-SM interaction via new scalar and vector
  mediators in rotating neutron stars}},
  \href{https://doi.org/10.1088/1475-7516/2021/09/027}{\emph{JCAP} {\bfseries
  09} (2021) 027} [\href{https://arxiv.org/abs/2106.10353}{{\ttfamily
  2106.10353}}].

\bibitem{Guha:2024pnn}
A.~Guha and D.~Sen, \emph{{Constraining the mass of fermionic dark matter from
  its feeble interaction with hadronic matter via dark mediators in neutron
  stars}}, {\emph{Phys. Rev. D} {\bfseries Accepted} (2024) arXiv:2401.14419}
  [\href{https://arxiv.org/abs/2401.14419}{{\ttfamily 2401.14419}}].

\bibitem{Shakeri:2022dwg}
S.~Shakeri and D.R.~Karkevandi, \emph{{Bosonic dark matter in light of the
  NICER precise mass-radius measurements}},
  \href{https://doi.org/10.1103/PhysRevD.109.043029}{\emph{Phys. Rev. D}
  {\bfseries 109} (2024) 043029}
  [\href{https://arxiv.org/abs/2210.17308}{{\ttfamily 2210.17308}}].

\bibitem{Karkevandi:2021ygv}
D.R.~Karkevandi, S.~Shakeri, V.~Sagun and O.~Ivanytskyi, \emph{{Bosonic dark
  matter in neutron stars and its effect on gravitational wave signal}},
  \href{https://doi.org/10.1103/PhysRevD.105.023001}{\emph{Phys. Rev. D}
  {\bfseries 105} (2022) 023001}
  [\href{https://arxiv.org/abs/2109.03801}{{\ttfamily 2109.03801}}].

\bibitem{Lenzi:2022ypb}
C.H.~Lenzi, M.~Dutra, O.~Louren\c{c}o, L.L.~Lopes and D.P.~Menezes, \emph{{Dark
  matter effects on hybrid star properties}},
  \href{https://doi.org/10.1140/epjc/s10052-023-11416-y}{\emph{Eur. Phys. J. C}
  {\bfseries 83} (2023) 266}
  [\href{https://arxiv.org/abs/2212.12615}{{\ttfamily 2212.12615}}].

\bibitem{Lourenco:2022fmf}
O.~Louren\c{c}o, C.H.~Lenzi, T.~Frederico and M.~Dutra, \emph{{Dark matter
  effects on tidal deformabilities and moment of inertia in a hadronic model
  with short-range correlations}},
  \href{https://doi.org/10.1103/PhysRevD.106.043010}{\emph{Phys. Rev. D}
  {\bfseries 106} (2022) 043010}
  [\href{https://arxiv.org/abs/2208.06067}{{\ttfamily 2208.06067}}].

\bibitem{Bell:2020lmm}
N.F.~Bell, G.~Busoni, S.~Robles and M.~Virgato, \emph{{Improved Treatment of
  Dark Matter Capture in Neutron Stars II: Leptonic Targets}},
  \href{https://doi.org/10.1088/1475-7516/2021/03/086}{\emph{JCAP} {\bfseries
  03} (2021) 086} [\href{https://arxiv.org/abs/2010.13257}{{\ttfamily
  2010.13257}}].

\bibitem{Tananbaum:2014oda}
H.~Tananbaum, M.C.~Weisskopf, W.~Tucker, B.~Wilkes and P.~Edmonds,
  \emph{{Highlights and Discoveries from the Chandra X-ray Observatory}},
  \href{https://doi.org/10.1088/0034-4885/77/6/066902}{\emph{Rept. Prog. Phys.}
  {\bfseries 77} (2014) 066902}
  [\href{https://arxiv.org/abs/1405.7847}{{\ttfamily 1405.7847}}].

\bibitem{Prinz:2015jkd}
T.~Prinz and W.~Becker, \emph{{A Search for X-ray Counterparts of Radio
  Pulsars}}, \href{https://doi.org/10.48550/arXiv.1511.07713}{\emph{Astrophys.
  J.} {\bfseries resubmitted} (2015) arXiv:1511.07713}
  [\href{https://arxiv.org/abs/1511.07713}{{\ttfamily 1511.07713}}].

\bibitem{Prinz:2015}
W.~{Becker}, \emph{{A Search for X-ray Counterparts of Radio Pulsars}},  in
  \emph{AAS/High Energy Astrophysics Division \#14}, vol.~14 of \emph{AAS/High
  Energy Astrophysics Division}, p.~114.08, Aug., 2014.

\bibitem{Sugahara:1993wz}
Y.~Sugahara and H.~Toki, \emph{{Relativistic mean field theory for unstable
  nuclei with nonlinear sigma and omega terms}},
  \href{https://doi.org/10.1016/0375-9474(94)90923-7}{\emph{Nucl. Phys. A}
  {\bfseries 579} (1994) 557}.

\bibitem{Glendenning:1991es}
N.K.~Glendenning and S.A.~Moszkowski, \emph{{Reconciliation of neutron star
  masses and binding of the lambda in hypernuclei}},
  \href{https://doi.org/10.1103/PhysRevLett.67.2414}{\emph{Phys. Rev. Lett.}
  {\bfseries 67} (1991) 2414}.

\bibitem{Typel:2009sy}
S.~Typel, G.~Ropke, T.~Klahn, D.~Blaschke and H.H.~Wolter, \emph{{Composition
  and thermodynamics of nuclear matter with light clusters}},
  \href{https://doi.org/10.1103/PhysRevC.81.015803}{\emph{Phys. Rev. C}
  {\bfseries 81} (2010) 015803}
  [\href{https://arxiv.org/abs/0908.2344}{{\ttfamily 0908.2344}}].

\bibitem{Taninah:2019cku}
A.~Taninah, S.E.~Agbemava, A.V.~Afanasjev and P.~Ring, \emph{{Parametric
  correlations in energy density functionals}},
  \href{https://doi.org/10.1016/j.physletb.2019.135065}{\emph{Phys. Lett. B}
  {\bfseries 800} (2020) 135065}
  [\href{https://arxiv.org/abs/1910.13007}{{\ttfamily 1910.13007}}].

\bibitem{Bhattacharjee:2013exa}
P.~Bhattacharjee, S.~Chaudhury and S.~Kundu, \emph{{Rotation Curve of the Milky
  Way out to $\sim$ 200 kpc}},
  \href{https://doi.org/10.1088/0004-637X/785/1/63}{\emph{Astrophys. J.}
  {\bfseries 785} (2014) 63} [\href{https://arxiv.org/abs/1310.2659}{{\ttfamily
  1310.2659}}].

\bibitem{Lu:2011wy}
B.-N.~Lu, E.-G.~Zhao and S.-G.~Zhou, \emph{{Quadrupole deformation
  $(\beta,\gamma)$ of light $\Lambda$ hypernuclei in constrained relativistic
  mean field model: shape evolution and shape polarization effect of $\Lambda$
  hyperon}}, \href{https://doi.org/10.1103/PhysRevC.84.014328}{\emph{Phys. Rev.
  C} {\bfseries 84} (2011) 014328}
  [\href{https://arxiv.org/abs/1104.4638}{{\ttfamily 1104.4638}}].

\bibitem{Xia:2022dvw}
C.-J.~Xia, T.~Maruyama, A.~Li, B.Y.~Sun, W.-H.~Long and Y.-X.~Zhang,
  \emph{{Unified neutron star EOSs and neutron star structures in RMF models}},
  \href{https://doi.org/10.1088/1572-9494/ac71fd}{\emph{Commun. Theor. Phys.}
  {\bfseries 74} (2022) 095303}
  [\href{https://arxiv.org/abs/2208.12893}{{\ttfamily 2208.12893}}].

\bibitem{Tolman:1939jz}
R.C.~Tolman, \emph{{Static solutions of Einstein's field equations for spheres
  of fluid}}, \href{https://doi.org/10.1103/PhysRev.55.364}{\emph{Phys. Rev.}
  {\bfseries 55} (1939) 364}.

\bibitem{Oppenheimer:1939ne}
J.R.~Oppenheimer and G.M.~Volkoff, \emph{{On Massive neutron cores}},
  \href{https://doi.org/10.1103/PhysRev.55.374}{\emph{Phys. Rev.} {\bfseries
  55} (1939) 374}.

\bibitem{Quaintrell:2003pn}
H.~Quaintrell, A.J.~Norton, T.D.C.~Ash, P.~Roche, B.~Willems, T.R.~Bedding
  et~al., \emph{{The mass of the neutron star in Vela X-1 and tidally induced
  non-radial oscillations in GP Vel}},
  \href{https://doi.org/10.1051/0004-6361:20030120}{\emph{Astron. Astrophys.}
  {\bfseries 401} (2003) 313}
  [\href{https://arxiv.org/abs/astro-ph/0301243}{{\ttfamily
  astro-ph/0301243}}].

\bibitem{Yakovlev:2004iq}
D.G.~Yakovlev and C.J.~Pethick, \emph{{Neutron star cooling}},
  \href{https://doi.org/10.1146/annurev.astro.42.053102.134013}{\emph{Ann. Rev.
  Astron. Astrophys.} {\bfseries 42} (2004) 169}
  [\href{https://arxiv.org/abs/astro-ph/0402143}{{\ttfamily
  astro-ph/0402143}}].

\bibitem{Navarro:1995iw}
J.F.~Navarro, C.S.~Frenk and S.D.M.~White, \emph{{The Structure of cold dark
  matter halos}}, \href{https://doi.org/10.1086/177173}{\emph{Astrophys. J.}
  {\bfseries 462} (1996) 563}
  [\href{https://arxiv.org/abs/astro-ph/9508025}{{\ttfamily
  astro-ph/9508025}}].

\bibitem{Bauer:2017qwy}
M.~Bauer and T.~Plehn, \emph{{Yet Another Introduction to Dark Matter}: {The
  Particle Physics Approach}}, vol.~959 of \emph{Lecture Notes in Physics},
  Springer (2019),
  \href{https://doi.org/10.1007/978-3-030-16234-4}{10.1007/978-3-030-16234-4},
  [\href{https://arxiv.org/abs/1705.01987}{{\ttfamily 1705.01987}}].

\bibitem{SENSEI:2020dpa}
{\scshape SENSEI} collaboration, \emph{{SENSEI: Direct-Detection Results on
  sub-GeV Dark Matter from a New Skipper-CCD}},
  \href{https://doi.org/10.1103/PhysRevLett.125.171802}{\emph{Phys. Rev. Lett.}
  {\bfseries 125} (2020) 171802}
  [\href{https://arxiv.org/abs/2004.11378}{{\ttfamily 2004.11378}}].

\bibitem{EDELWEISS:2020fxc}
{\scshape EDELWEISS} collaboration, \emph{{First germanium-based constraints on
  sub-MeV Dark Matter with the EDELWEISS experiment}},
  \href{https://doi.org/10.1103/PhysRevLett.125.141301}{\emph{Phys. Rev. Lett.}
  {\bfseries 125} (2020) 141301}
  [\href{https://arxiv.org/abs/2003.01046}{{\ttfamily 2003.01046}}].

\bibitem{PandaX-II:2021nsg}
{\scshape PandaX-II} collaboration, \emph{{Search for Light Dark
  Matter-Electron Scatterings in the PandaX-II Experiment}},
  \href{https://doi.org/10.1103/PhysRevLett.126.211803}{\emph{Phys. Rev. Lett.}
  {\bfseries 126} (2021) 211803}
  [\href{https://arxiv.org/abs/2101.07479}{{\ttfamily 2101.07479}}].

\bibitem{XENON:2019gfn}
{\scshape XENON} collaboration, \emph{{Light Dark Matter Search with Ionization
  Signals in XENON1T}},
  \href{https://doi.org/10.1103/PhysRevLett.123.251801}{\emph{Phys. Rev. Lett.}
  {\bfseries 123} (2019) 251801}
  [\href{https://arxiv.org/abs/1907.11485}{{\ttfamily 1907.11485}}].

\bibitem{SuperCDMS:2020ymb}
{\scshape SuperCDMS} collaboration, \emph{{Constraints on low-mass, relic dark
  matter candidates from a surface-operated SuperCDMS single-charge sensitive
  detector}}, \href{https://doi.org/10.1103/PhysRevD.102.091101}{\emph{Phys.
  Rev. D} {\bfseries 102} (2020) 091101}
  [\href{https://arxiv.org/abs/2005.14067}{{\ttfamily 2005.14067}}].

\bibitem{DAMIC:2019dcn}
{\scshape DAMIC} collaboration, \emph{{Constraints on Light Dark Matter
  Particles Interacting with Electrons from DAMIC at SNOLAB}},
  \href{https://doi.org/10.1103/PhysRevLett.123.181802}{\emph{Phys. Rev. Lett.}
  {\bfseries 123} (2019) 181802}
  [\href{https://arxiv.org/abs/1907.12628}{{\ttfamily 1907.12628}}].

\end{thebibliography}\endgroup
\bibliographystyle{JHEP}

\end{document}